\documentclass[review]{elsarticle}

\usepackage{hyperref}
\usepackage{graphicx}
\usepackage{multirow}
\usepackage{pbox}
\usepackage{siunitx}

\journal{Journal of Ultramicroscopy}
\makeatletter
\def\ps@pprintTitle{%
 \def\@oddhead{\small \copyright~2016. This manuscript version is made available under the \href{https://creativecommons.org/licenses/by-nc-nd/4.0/}{CC-BY-NC-ND 4.0 license}}
 \let\@oddhead\@oddhead
 \let\@evenhead\@empty
 \def\@oddfoot{\small Published in Ultramicroscopy 169 (2016) 98-106. DOI: \href{http://dx.doi.org/10.1016/j.ultramic.2016.07.004}{10.1016/j.ultramic.2016.07.004}}
 \let\@evenfoot\@oddfoot}
\makeatother

\begin{document}

\begin{frontmatter}

\title{Assessing electron beam sensitivity for SrTiO$_3$ and La$_{0.7}$Sr$_{0.3}$MnO$_3$ using Electron Energy Loss Spectroscopy}

\author[NTNU_IFY]{Magnus Nord\corref{correspondingauthor}\fnref{presentaddress}}
\author[NTNU_IFY,SINTEF]{Per Erik Vullum}
\author[NTNU_IET]{Ingrid Hallsteinsen}
\author[NTNU_IET]{Thomas Tybell}
\author[NTNU_IFY]{Randi Holmestad}

\address[NTNU_IFY]{Department of Physics, NTNU, Trondheim, Norway}
\address[NTNU_IET]{Department of Electronics and Telecommunications, NTNU, Trondheim, Norway}
\address[SINTEF]{Materials and Chemistry, SINTEF, Trondheim, Norway}

\cortext[correspondingauthor]{Corresponding author}
\fntext[presentaddress]{Present address: SUPA, School of Physics and Astronomy, University of Glasgow, Glasgow G12 8QQ, United Kingdom}

\begin{abstract}
    Thresholds for beam damage have been assessed for La$_{0.7}$Sr$_{0.3}$MnO$_3$ and SrTiO$_3$ as a function of electron probe current and exposure time at 80 and 200 kV acceleration voltage.
    The materials were exposed to an intense electron probe by aberration corrected scanning transmission electron microscopy (STEM) with simultaneous acquisition of electron energy loss spectroscopy (EELS) data.
    Electron beam damage was identified by changes of the core loss fine structure after quantification by a refined and improved model based approach.
    At 200 kV acceleration voltage, damage in SrTiO$_3$ was identified by changes both in the EEL fine structure and by contrast changes in the STEM images. 
    However, the changes in the STEM image contrast as introduced by minor damage can be difficult to detect under several common experimental conditions. 
    No damage was observed in SrTiO$_3$ at 80 kV acceleration voltage, independent of probe current and exposure time.
    In La$_{0.7}$Sr$_{0.3}$MnO$_3$, beam damage was observed at both 80 and 200 kV acceleration voltages.
    This damage was observed by large changes in the EEL fine structure, but not by any detectable changes in the STEM images.
    The typical method to validate if damage has been introduced during acquisitions is to compare STEM images prior to and after spectroscopy.
    Quantifications in this work show that this method possibly can result in misinterpretation of beam damage as changes of material properties.
\end{abstract}


\end{frontmatter}

\section{Introduction}
Perovskite oxide materials have received a great deal of interest due to their magnetic and electronic properties\cite{mannhart_oxide_2010,hammerl_shedding_2011}.
In bulk form they exhibit a wide range of functional properties, such as ferroelectricity\cite{martin_advances_2010}, ferromagnetism\cite{dagotto_colossal_2001}, dielectric properties, and colossal magnetoresistance\cite{dagotto_colossal_2001}.
This range of properties are enabled by strong structure-function coupling, where small variations in structural parameters can result in large changes in functional response.
In recent years there has been a resurgence of interest in these materials due to advances in thin film synthesis methods such as molecular beam epitaxy and pulsed laser deposition (PLD), where epitaxial growth can be controlled down to single monolayers\cite{martin_advances_2010}.
By epitaxial growth and correct substrate use it is possible to control the structure, opening for new avenues for fine tuning functional properties.

One such control parameter is chemical substitution of the A- or B-cations. 
For example, by replacing La with Sr in La$_{1-x}$Sr$_{x}$MnO$_3$ the magnetic response can be tuned, and at x $\approx$ 0.3 maximum in colossal magnetoresistance is observed\cite{lee_controlling_2012}.
Strain engineering is another control parameter, by changing the in-plane lattice constant of the substrate the thin film's in-plane lattice spacings can be locked to the substrate's.
This clamping can modify the crystal structure of the film, for example through biaxial strain and suppression of oxygen octahedral rotations\cite{biegalski_impact_2015,he_towards_2015}.
Oxygen vacancies constitute an important control parameter.
It has been reported that by growing oxygen deficient La$_{0.5}$Sr$_{0.5}$CoO$_{3-\delta}$ on substrates with different lattice parameters, the oxygen vacancies can order in specific crystal directions\cite{gazquez_lattice_2013}.
Oxygen vacancies can also affect the conductivity of a material, for example turning the insulating SrTiO$_3$ into a conductor through charge transfer from the vacancy to the titanium atom\cite{basletic_mapping_2008}.
In La$_{0.7}$Sr$_{0.3}$MnO$_3$, the presence of oxygen vacancies can lead to breakdown of the ferromagnetic order by suppressing the double-exchange mechanism\cite{khartsev_colossal_2000}.
Lastly, crystal orientation of the substrate adds another control parameter.
As materials often are anisotropic, growing films in different crystal orientations is paramount. 
The most studied substrate orientation has been the (100) orientation, however recently (111) oriented thin film systems have been realized\cite{hallsteinsen_surface_2013,oshea_nanoscale_2015}.
By relying on the discussed control parameters, it should be possible to fine tune functional properties and to tailor-make devices.
However, the complex interplay between the different parameters makes it challenging to understand and characterize the relationship between structure and properties, generating a demand for high resolution spatial techniques.

Transmission electron microscopy (TEM) is one of the most used tools to study perovskite oxides\cite{oshea_nanoscale_2015,tan_2d_2011,kimoto_element-selective_2007,nakagawa_why_2006,muller_atomic-scale_2004,yao_electron-beam-induced_2014,he_towards_2015}, and in particular embedded parts of the materials. 
In the latter years, the development of sub-\AA~resolution scanning TEM\cite{haider_spherical-aberration-corrected_1998,nellist_direct_2004} (STEM) combined with high energy resolution electron energy loss spectroscopy (EELS) has turned the combination of STEM and EELS into one of the most powerful tools to characterize the structure and electronic properties of perovskite oxides with atomic resolution\cite{tan_2d_2011,kimoto_element-selective_2007}.
EEL spectra are now used to collect information about chemical composition including cation diffusion\cite{sankara_rama_krishnan_misfit_2014}, cation oxidation state\cite{varela_atomic-resolution_2009,muller_atomic-scale_2008}, crystal structure modifications\cite{nishida_effect_2013}, and vacancies\cite{nakagawa_why_2006,muller_atomic-scale_2004}.
These are all parameters that the functional properties are extremely sensitive to.
However, a correct interpretation of structure-property relations relies on several sensitive steps in the characterization scheme: a) the high energy and flux of the electron beam used in modern TEMs can possibly modify the materials due to beam damage\cite{egerton_radiation_2004,yao_electron-beam-induced_2014,garvie_electron-beam-induced_1994,houben_spatial_2012,botton_elemental_2010}, b) the interpretation of the fine structure information of the EEL core loss spectra rely on a correct handling of the plasmon background and modelling of the various features in the spectra, and c) the TEM sample preparation can possibly alter the inherent structure and properties of the materials.

In the present paper we have used STEM-EELS to systematically study an epitaxial heterostructure with two of the most studied perovskite oxides: La$_{0.7}$Sr$_{0.3}$MnO$_3$ (henceforth LSMO) grown on SrTiO$_3$ (STO) along the {[}111{]} direction. 
A refined and improved model based approach for processing of EELS spectra in order to correctly interpret the fine structure of the core loss is presented.
From this we have established thresholds for beam current and exposure times, beyond which beam damage is introduced. 
One very important observation is that for LSMO the onset of beam damage, both as a function of beam current and exposure time, sets in before any damage can be seen directly in the STEM image. 
For STO, changes in the STEM image occur simultaneously with the onset of beam damage as deduced from EELS, however some of these subtle changes can be missed under common experimental conditions.

\section{Materials and method}
\subsection{Experiment}\label{sec:experiment}

\begin{figure}[hb]
  \centering
    \includegraphics[width=1.0\columnwidth]{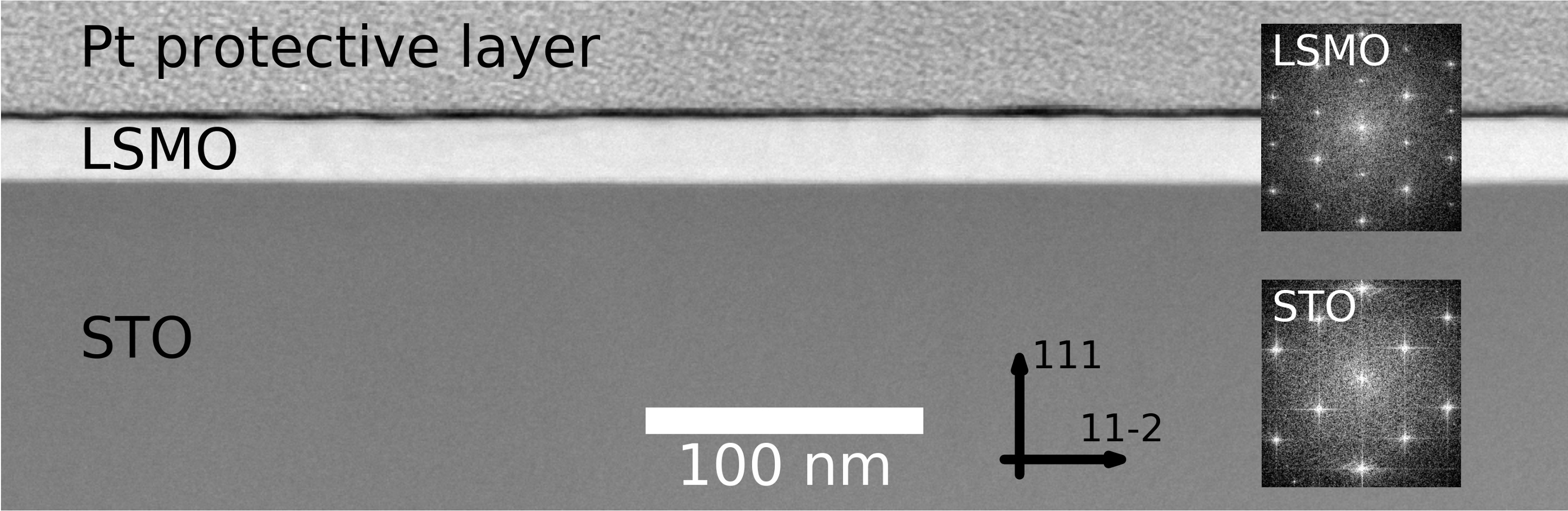}
  \caption
  {HAADF STEM overview of the LSMO film, STO substrate and the FIB deposited Pt protection layer.
  Insets show FFTs from HRTEM images acquired on the same sample.}
  \label{fig:stem_fft_overview}
\end{figure}

The beam exposure experiments were performed on LSMO/STO:Nb-(111) and LSMO/LaFeO$_3$ (LFO)/STO-(111) samples grown by PLD\cite{hallsteinsen_surface_2013}.
Cross section TEM lamellas were prepared by Focused Ion Beam (FIB), on a FEI Helios Nanolab DualBeam FIB, using standard lift-out technique. 
Prior to starting the FIB preparation, a 10 nm Pt/Pd layer was sputter coated on top of the wafer.
In the FIB an additional 80 nm Pt protection layer was deposited by electron beam assisted deposition, before adding a \SI{3}{\micro\meter} carbon protection layer by ion beam assisted deposition.
The coarse ion beam thinning was done at 30 kV. 
Final thinning was performed with 5 and 2 kV ion beam acceleration voltages.
After FIB preparation, the TEM samples were milled for 20 seconds on each side with Ar-ions at 100 eV using a Gatan PIPS II.
The TEM experiments were done on a double-corrected Jeol ARM200CF equipped with a Gatan Quantum ER, using an energy dispersion of 0.25 eV/channel and a collection semi-angle of 66 mrad.
Low and core loss EEL spectra were acquired quasi-simultaneously using the DualEELS functionality on the Gatan Quantum. 
All STEM-high angle annular dark field (HAADF) data were acquired with inner and outer collection angles of 118-471 mrad.
A HAADF-STEM overview of the sample is shown in Fig. \ref{fig:stem_fft_overview}.

The beam exposure experiments were performed by acquiring a STEM-HAADF image of the area, then positioning the STEM probe inside this area while simultaneously acquiring an EEL spectrum every 0.1 seconds. 
The probe remained at this position for up to 3.3 minutes.
After the exposure experiment was finished, another STEM-HAADF image was acquired for comparison and to estimate the sample drift.
The sample drift was estimated by using some landmark features in the field of view, for example an inhomogeneity in the Pt-protective layer for the experiments done on the LSMO film.
For the STO substrate, which has no inherent identifiable features, a marker was created by leaving the electron probe for a sufficient time to change the HAADF intensity, like in Fig. \ref{fig:sto_stem_finestructure_line_scan}.
This procedure was repeated for different acceleration voltages, spot sizes and condenser apertures on both the LSMO film, and the STO substrate. 
The properties of the different probes are shown in Table \ref{table:probe_properties}.
For the LSMO film, all the exposure experiments were done in the middle of the cross section film: about 10 nm from the interface.
The STO substrate exposures were performed about 100 nm away from the interface.
The TEM foil thickness in the analysed regions was approximately 0.5 $\lambda$ ($\lambda =$ inelastic mean free path) at 200 kV, and 0.6$\lambda$ at 80 kV.

\subsection{EELS modelling}\label{sec:eels_processing_method}
\begin{figure}[hb]
  \centering
    \includegraphics[width=1.0\columnwidth]{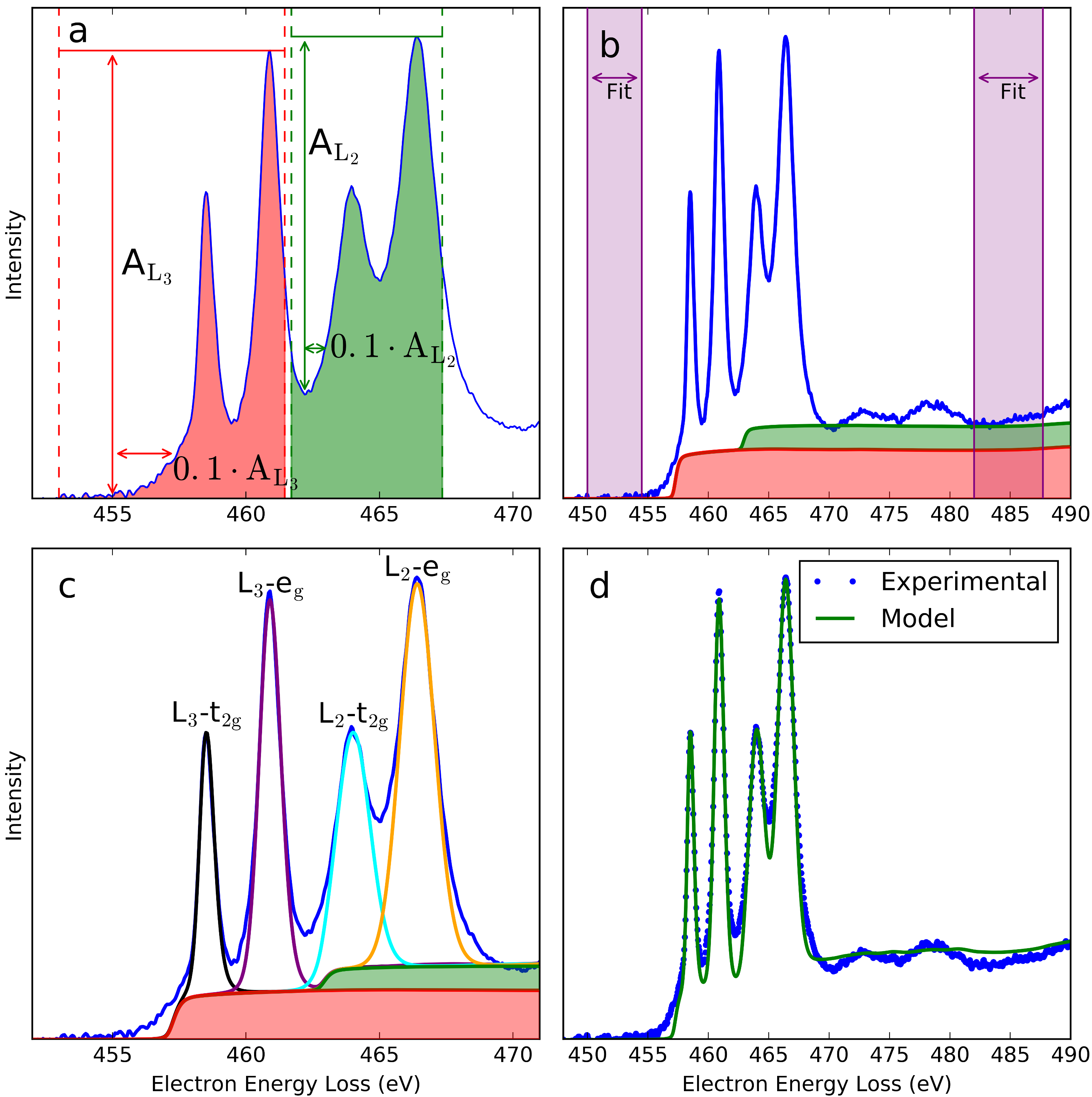}
  \caption
  {Modelling of SrTiO$_3$ Ti-L$_{2,3}$ white lines. a) Showing how the onset of the Ti-L$_3$ and L$_2$ Hartree-Slater core loss edges are calculated. 
      The minimum and maximum intensities are found in the red (L$_3$) area, which gives the amplitude A$_{\mathrm{L}_3}$. 
      This amplitude times 0.1 gives the onset intensity, which is used to find the L$_3$ onset energy. 
      The L$_2$ onset is found in the same way, using the green area.
      b) Purple areas show the pre- and post-edge fitting areas for the intensity of the Hartree-Slater core loss edges. 
      The L$_2$ Hartree-Slater edge intensity is locked to half the L$_3$ Hartree-Slater edge intensity.
      The resulting core loss edges are shown with the green and red areas. 
      c) The Ti-L$_{2,3}$ white lines fitted using four Gaussian distributions.
      d) The full fitted model, which is a sum of all the components shown in c).
  }
  \label{fig:til23_eels_modelling}
\end{figure}

To extract physical relevant parameters from the EELS fine structure data, the model based approach was utilized\cite{manoubi_curve_1990} using the open source software HyperSpy\cite{francisco_de_la_pena_hyperspy:_2015}.
This approach works by fitting several components to an experimental spectrum, and the sum of these components is the model. 
These components are distributions or functions such as Gaussians or Hartree-Slater core loss ionization edges\cite{ahn_inner_1985}. 
For modelling the Titanium L$_{2,3}$ core loss edge (transitions between 2p$^{1/2}$ and 2p$^{3/2}$ to 3d), four Gaussians were used (one for each peak), and two Hartree-Slater edges which model the L$_2$ and L$_3$ ionization edges (Fig. \ref{fig:til23_eels_modelling}b).
The Manganese L$_{2,3}$ peaks were modelled in the same fashion, but only by using two Gaussians and two Hartree-Slater edges.
The Oxygen-K edge was modelled using three Gaussians.
The core loss ionization edge of the O-K edge was not modelled directly using its own component, due to the difficulty of setting a robust edge onset energy.
However the changes in the fine structure were still picked up in a robust fashion using the Gaussians.
For all core loss edges, one could use different components (such as Voigt functions) or add more Gaussians to potentially get better models.
However, this would increase the amount of free variables, causing less robust fitting, which again could lead to misinterpretation of data.
Thus the amount of components for each core loss edge was chosen so that, a) they would accurately model the changes of the EELS fine structure, b) they would fit the data robustly.

One advantage of the model based approach, is that the low loss plasmon signal is convolved with the core loss signal to account for multiple scattering.
Hence, possible artifacts introduced by more common deconvolution techniques are avoided. 
In addition, this will result in better fits, given that a suitable model of the core loss edges can be constructed.
The steps for fitting the Titanium L$_{2,3}$ EELS fine structure are outlined below, similar to the process used by Tan \emph{et al.}\cite{tan_oxidation_2012} to calculate the edge onset energy:
\begin{enumerate}
    \item Calibrate the energy offset by using the zero loss peak (ZLP), which is acquired quasi-simultaneously.
    \item Do principal component analysis (PCA) for increasing the signal-to-noise ratio\cite{varela_atomic-resolution_2009}.
    \item Fit the power law background and freeze it.
    \item Determine the Ti-L$_3$ ionization edge onset energy (Fig. \ref{fig:til23_eels_modelling}a).
    \begin{enumerate}
        \item The edge onset energy is set to a percentage of the net height of the L$_3$ peak.
        \item If this percentage value is set too low, the edge onset energy can be influenced by pre-edge noise. But if it is too high, the edge onset energy can be affected by the shape of the fine structure. For the present datasets, 10\% of the net height of the peak was chosen, in-line with Tan \emph{et al}.\cite{tan_oxidation_2012}.
    \end{enumerate}
    \item The Ti-L$_2$ edge onset energy is set the same way as Ti-L$_3$, where the net height of the L$_2$ peak is calculated by using the lowest point between the L$_{2,3}$ peaks, and the highest point in the L$_2$ peaks (Fig. \ref{fig:til23_eels_modelling}a).
    \item Lock the edge onset energy for both Ti-L$_2$ and L$_3$.
    \item Fit ionization edges to pre- and post-edge area (Fig. \ref{fig:til23_eels_modelling}b, purple regions). The L$_2$ ionization edge (green region) intensity is set to half the intensity of the L$_3$\cite{ahn_inner_1985} ionization edge (red region), due to that the 2p$^{3/2}$ and 2p$^{1/2}$ electron orbitals have four and two electrons, respectively.
    \item Use four Gaussians to fit the white lines: firstly fitting each Gaussian in a narrow region around each peak, starting with the most intense (e$_g$), then fitting all the Gaussians simultaneously without any constraints.
\end{enumerate}

An important consideration is the effect of mixed oxidation states on this method.
For example, the Mn-L$_{2,3}$ EELS fine structure for a mix of 50\% Mn$^{+2}$ and 50\% Mn$^{+4}$ would be different compared to 100\% Mn$^{+3}$.
The former gives a signal that is a superposition of the spectra from Mn$^{+2}$ and Mn$^{+4}$. 
This mixed Mn$^{+2}$ and Mn$^{+4}$ spectrum (with an average oxidation state of +3) is significantly different from the Mn$^{+3}$ spectrum. 
Due to chemical shift the Mn$^{+2}$ spectrum would be at lower energy loss compared to the Mn$^{+4}$, while the Mn$^{+3}$ would be somewhere in the middle.
Thus, the mixed oxidation state spectrum peak width would be larger than the single oxidation state spectrum, which would be detectable through the sigma value of the L$_3$ and L$_2$ Gaussians\cite{egerton_electron_2009}.
In co-junction with the Mn-L$_3$ core loss ionization edge onset\cite{tan_oxidation_2012}, this could be a robust way of detecting mixed oxidation state systems.
For the Mn-L$_{2,3}$ datasets analyzed in this work there was the opposite effect, a narrowing of the EELS fine structure.

\subsection{Assessing beam damage}\label{sec:assessing_beam_damage}
Using the parameters from the Gaussians discussed above, one can calculate attributes.
These attributes are values like the Ti-e$_g$/t$_{2g}$ ratio and Mn-L$_{2,3}$ energy separation, which can be used to calculate physical properties like oxidation states.
To find when beam damage occurs in the material, one can find the point where the change in an attribute is significant, i.e. exceeds random variations.
The beginning of the beam exposure is used as reference to set an initial value for the attribute for an undamaged region.
For example between 0-10 nC/nm$^2$ in Fig. \ref{fig:sto_200kv_beam_exposure}c.
For the datasets analyzed in this work, the random noise in the attributes were relatively high (see Fig. \ref{fig:sto_200kv_beam_exposure}c, blue transparent line) so the datasets were smoothed using a Gaussian blur (see Fig. \ref{fig:sto_200kv_beam_exposure}c, orange line).
Next, a calibration dataset with very low electron beam current pr. area was used to calculate the uncertainty for each of the attributes, these values are shown in Tab. 1 in the supporting information.

To know if the change of an attribute translates into a significant physical change is important. 
However, it is outside the scope of this work to find the oxidation states of Mn and Ti as a function of beam exposure.
In addition, having a grasp on the uncertainty is important for knowing how sensitive the method is in detecting changes in the material.
The sensitivity ranged from about 0.05 to 0.18 oxidation state, which makes it possible to detect physically relevant changes.
The details of this are described in the supporting information.

Converting the attributes to physical properties can be done using literature values of materials with known oxidation states.
For Titanium, the oxidation state is determined by comparing the Ti-L$_{2,3}$ t$_{2g}$/e$_g$ intensity ratio or the average center position of the four Ti-L$_{2,3}$ peaks with materials where the Ti-oxidation state is known.
Here, spectra from BaTi$_{1-x}$Nb$_x$O$_3$ by Shao \emph{et al}.\cite{shao_quantification_2010} were used to get ballpark estimates of the relation between the t$_{2g}$/e$_g$ intensity ratio and the oxidation state.
The intensity ratio and the center difference between the O-K prepeak and peak B were also analyzed.
The O peaks were less sensitive than the Ti-L$_{2,3}$ peaks to beam damage, and they did not show any significant changes as a function of changing Ti oxidation state.
Similar values calculated for Manganese is shown in the supporting information.

\section{Results and discussion}

\begin{figure}[hb]
  \centering
    \includegraphics[width=1.0\columnwidth]{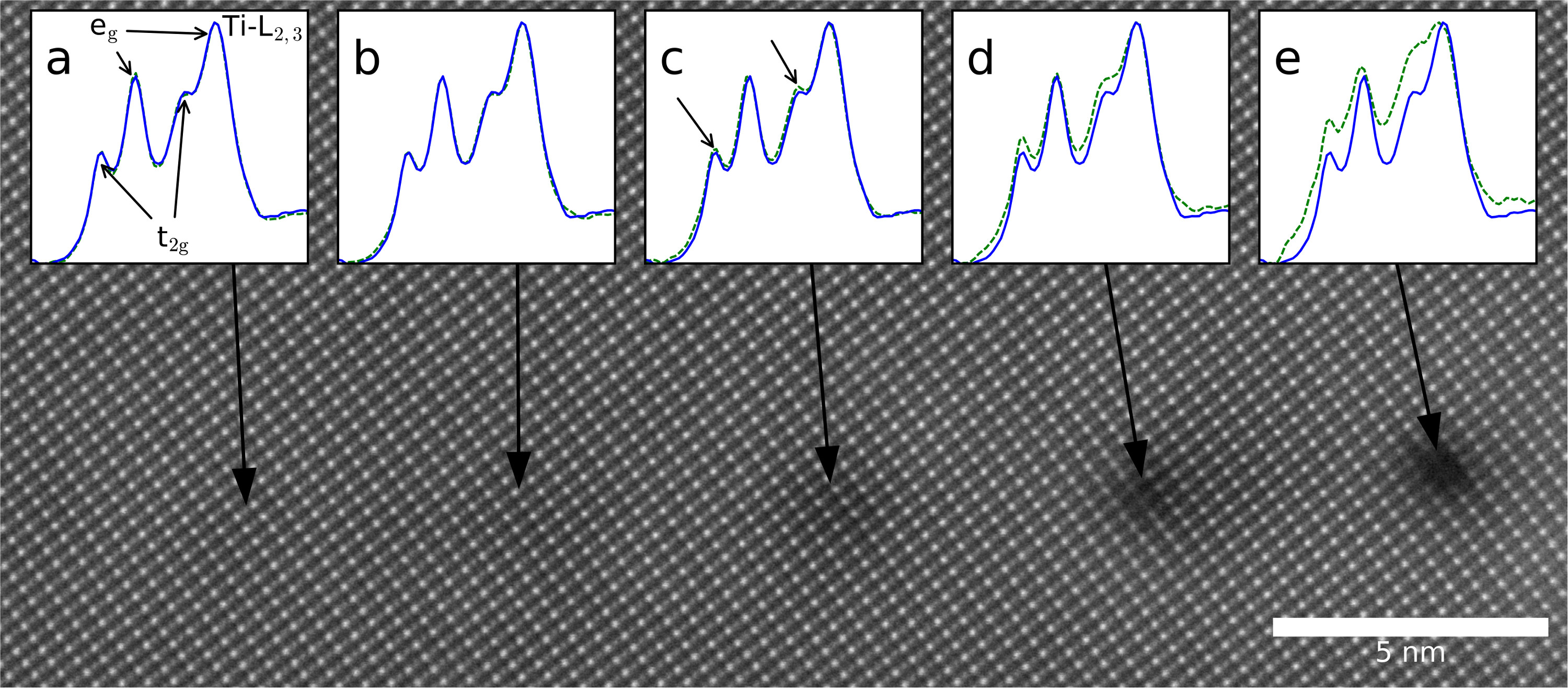}
  \caption{
    STEM-HAADF image of STO after beam exposure at 200 kV in the points highlighted by the arrows.
    The leftmost area a) having received the least amount of electron dose, and the rightmost  e) the most.
    After the beam exposures, a line scan was done across all the exposed points.
    Insets show the Ti-L$_{2,3}$ edge from an unexposed area (solid blue) and the exposed points (dashed green).
    The insets show an increasing amount of sample damage with increasing electron dose, as expected.
    No changes in HAADF intensity and the Ti-L$_{2,3}$ edge are observed in a) and b).
    In c) there are some subtle changes, indicated by the arrows, in the form of increasing intensity in the t$_{2g}$ peaks, and a small change in the HAADF intensity.
d) and e) show larger changes in both the Ti-L$_{2,3}$ and the HAADF intensity.
  }
  \label{fig:sto_stem_finestructure_line_scan}
\end{figure}

To assess the sensitivity of STO to electron beam exposure, several locations in vicinity to each other were exposed to different electron doses. 
An EELS line scan was acquired across these exposed locations to systematically quantify the EEL spectra.
The result of such an experiment at 200 kV is seen in Fig. \ref{fig:sto_stem_finestructure_line_scan}.
The HAADF STEM image has five exposed locations with increasing dose going from the left to the right.
Dashed green lines are the signal from the exposed regions, and the solid blue line is the reference from a location that was not exposed prior to the line scan.
For the two least exposed locations, a) and b), there are no detectable changes in the EELS fine structure of Ti-L$_{2,3}$.
For c) there are some subtle changes highlighted with the arrows: the e$_g$-peaks increase slightly compared to the t$_{2g}$-peaks.
In d) and e) there are significant changes in the EELS fine structure.
These changes match well with the corresponding changes in the HAADF signal intensity.
This is consistent with a removal of atoms through electron-beam sputtering\cite{egerton_radiation_2004}, which leads to a thinner TEM foil. 
Electron beam sputtering is further confirmed by a strong decrease of the plasmon peak signal, seen in Fig. 2 in the supporting information.
Even for c), with subtle EELS fine structure changes, there is a slight decrease in the HAADF intensity.
Hence, for STO a change in the HAADF intensity can be used to indicate beam damage.
In Fig. \ref{fig:sto_stem_finestructure_line_scan} the entire dynamic range of the STEM-detector is used to visualize the contrast of STO.
If other phases than STO are present in the STEM image (for instance vacuum or Pt), only a limited part of the detector's dynamic range will be used to display the contrast variations in STO, which can possibly wipe out contrast changes due to beam damage.

\begin{figure}[hb]
  \centering
    \includegraphics[width=1.0\columnwidth]{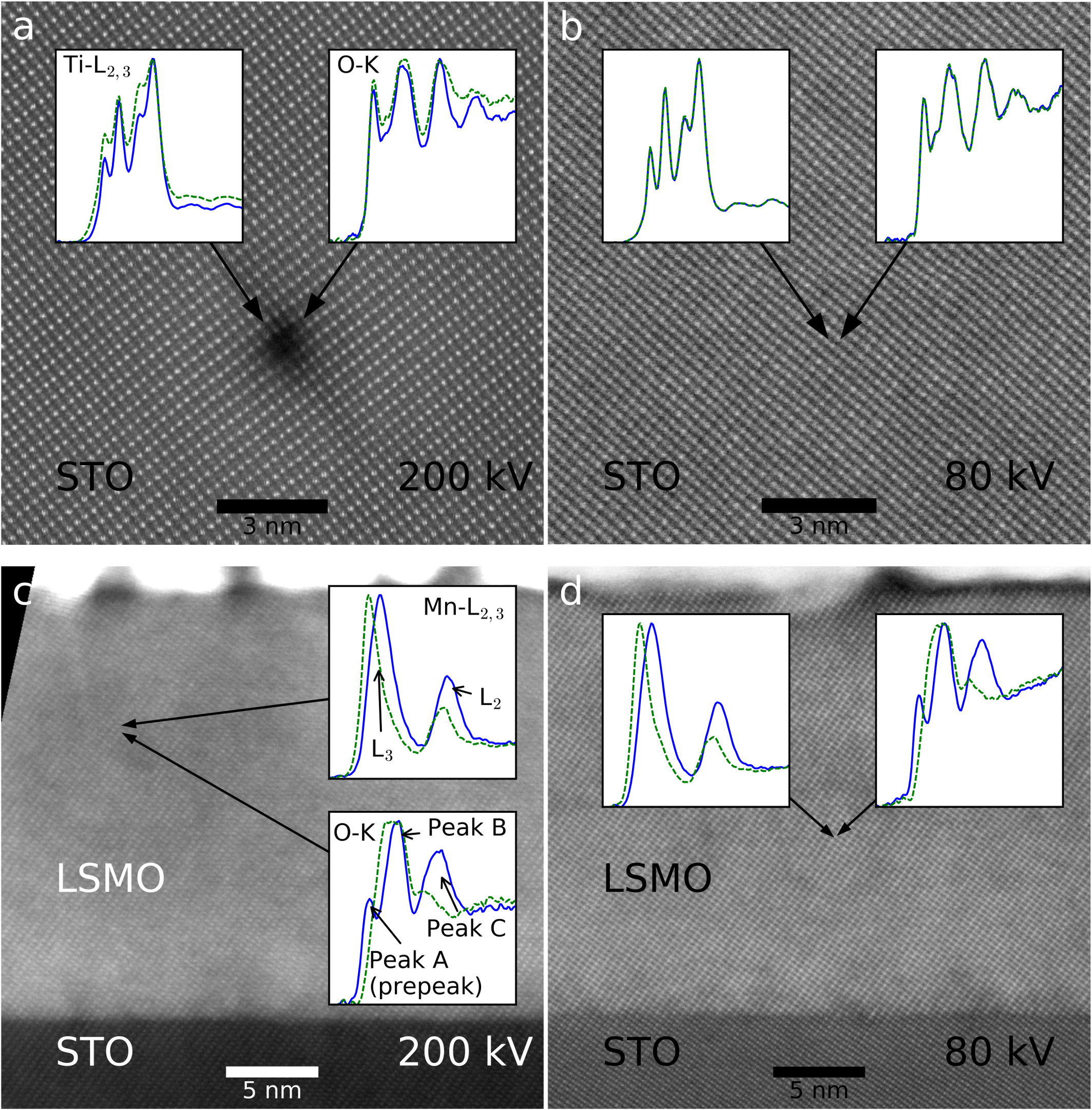}
  \caption{
      STEM-HAADF images acquired after a long beam exposure, arrows highlighting the exposed points. 
      Insets showing the B-cation L$_{2,3}$ and oxygen-K core loss EELS, before (solid blue) and after (dashed green) the beam exposure.
      a) STO after exposure to a 200 kV electron beam, shows changes in the HAADF intensity and EELS fine structure.
      b) STO after exposure to a 80 kV electron beam, showing no changes in HAADF and EELS.
      c) and d) LSMO after exposure to 200 and 80 kV electron beam, show no changes in HAADF intensity, but large changes in both EELS fine structures. 
      Arrows in the insets in c) highlight features in the EELS fine structure.
  }
  \label{fig:stem_fine_structure_exposure}
\end{figure}

Beam damage was further studied as a function of acceleration voltage, and by also including the LSMO thin film.
A combination of HAADF STEM-images and EEL spectra at both 80 and 200 kV for STO and LSMO is shown in Fig. \ref{fig:stem_fine_structure_exposure}.
The exposed areas are highlighted with arrows.
The insets show the EELS fine structure before (solid blue) and after (dashed green) the beam exposure experiment, for the B-cation (Ti-L$_{2,3}$ or Mn-L$_{2,3}$) and the O-K.
As expected from Fig. \ref{fig:sto_stem_finestructure_line_scan}, for the STO at 200 kV there is a clear reduction of HAADF intensity in co-junction with changes in the EELS fine structure.
At 80 kV no changes were observed in either the HAADF intensity nor the EELS fine structure for STO, independent of probe intensities and exposure time.
Electron beam induced damage in LSMO behaves very differently from beam damage in STO.
This is demonstrated at 200 kV in Fig. \ref{fig:stem_fine_structure_exposure}c, where large changes in the Mn-L$_{2,3}$ and O-K edges were observed despite that no changes could be seen in the corresponding HAADF-images.
The changes in the Mn-L$_{2,3}$, signified with an increase of the L$_{2,3}$-ratio, are consistent with a reduction of the Mn oxidation state\cite{varela_atomic-resolution_2009,tan_oxidation_2012}.
For the O-K edge, a decrease of both the prepeak and peak C and an increase of peak B indicate that oxygen vacancies are created\cite{yao_electron-beam-induced_2014}.
All these changes are consistent with oxygen being removed from the material:
oxygen vacancies lead to less charge transfer from manganese, giving a lower Mn oxidation state.
However, the low atomic number of oxygen compared to the cations makes the significant beam damage demonstrated in LSMO practically impossible to detect directly in the HAADF STEM images.

\begin{figure}[hb]
  \centering
    \includegraphics[width=1.0\columnwidth]{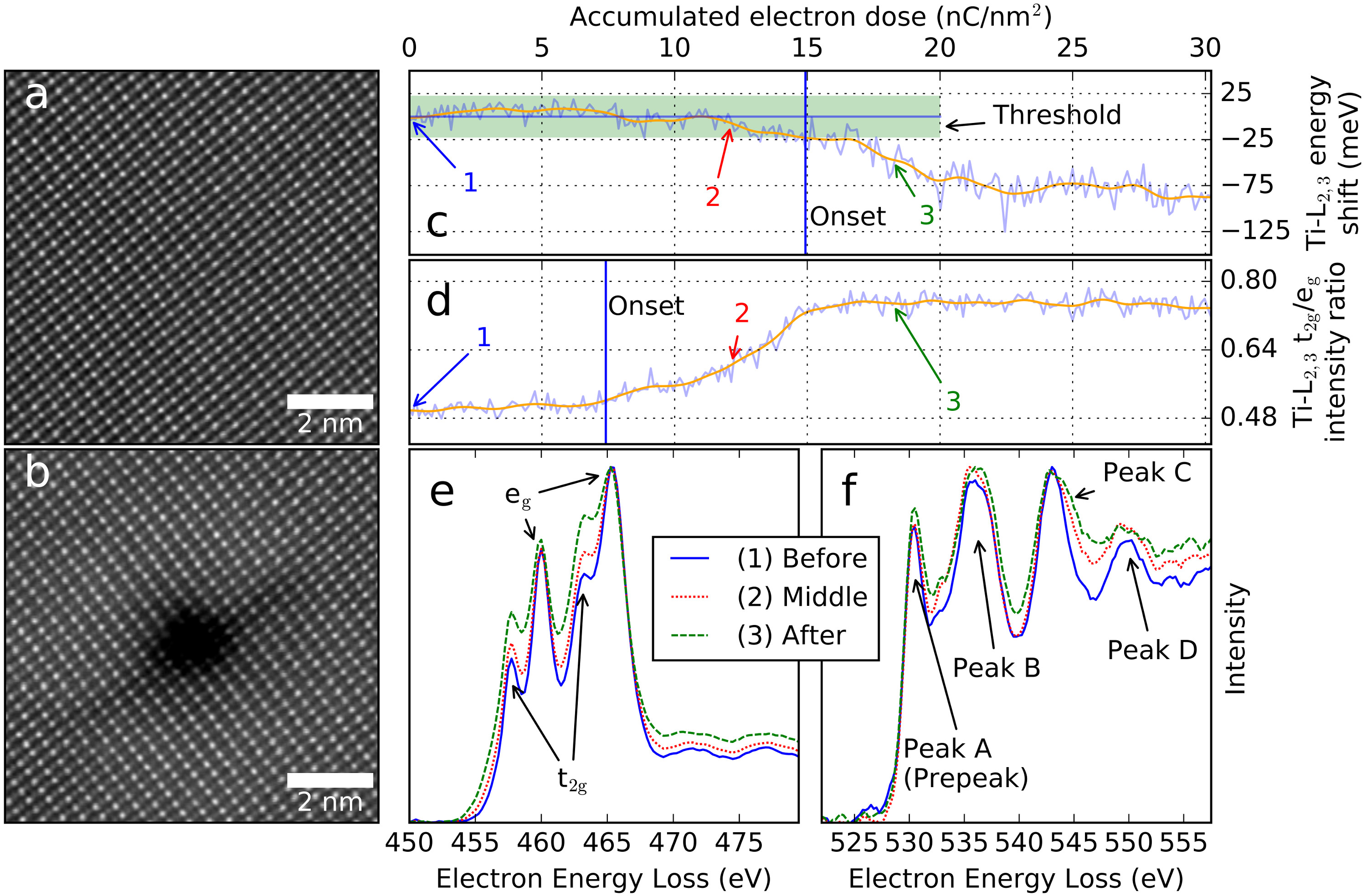}
  \caption
   {Beam exposure experiment on SrTiO$_3$ at 200 kV acceleration voltage. 
        a) STEM-HAADF image of SrTiO$_3$, before the experiment.
        b) Same as a), but after the experiment.
        c) Total energy shift of the Ti-L$_{2,3}$ white lines as a function of accumulated electron dose.
        The mean initial value is shown with the horizontal blue line, and the threshold is marked with the green transparent box. 
        The onset is set where the smoothed attribute (orange line) goes outside the threshold.
        d) Ratio of the Ti-L$_{2,3}$ e$_g$ and t$_{2g}$, as a function of accumulated electron dose.
        e) EEL spectra from the Titanium-L$_{2,3}$ core loss edge, showing the changes in the EELS fine structure before, during and after the beam exposure marked with the arrows in c) and d).
        f) Same as e), but for the O-K core loss edge. 
        For this specific probe current, none of the measured attributes in the O-K edge was significant. 
        However, there are clear changes in the EELS fine structure signified by a reduction of peak D.}
    \label{fig:sto_200kv_beam_exposure}
\end{figure}

Continuous acquisition of core loss EEL spectra during accumulated beam exposure was used to determine beam damage as a function of electron dose.
Such an experiment on STO at 200 kV is shown in Fig. \ref{fig:sto_200kv_beam_exposure}, where the beam has exposed the same point for 3.3 minutes (a total of 30 nC/nm$^2$).
Figs. \ref{fig:sto_200kv_beam_exposure}a and b show HAADF STEM images before and after the beam exposure experiment.
The EELS data processing method as explained in Sec. \ref{sec:eels_processing_method} was used to quantify the features in Ti-L$_{2,3}$ and O-K core loss edges.
The results are shown in Figs. \ref{fig:sto_200kv_beam_exposure}c and d.
The chemical shift of the Ti-L$_{2,3}$ peak as a function of the accumulated electron dose is shown in Fig. \ref{fig:sto_200kv_beam_exposure}c.
No chemical shift is observed until about 14 nC/nm$^2$, and at higher dose there is a shift towards lower energy.
This chemical shift is consistent with a change from Ti$^{+4}$ towards Ti$^{+3}$\cite{muller_atomic-scale_2004}.
In Fig. \ref{fig:sto_200kv_beam_exposure}d the intensity ratio between the Ti-L$_{2,3}$ e$_g$ and t$_{2g}$ peaks (highlighted in Fig. \ref{fig:sto_200kv_beam_exposure}e) is plotted.
From about 7 nC/nm$^2$ electron dose there is a clear increase in the t$_{2g}$/e$_g$ ratio, consistent with a decrease in the Ti oxidation state\cite{muller_atomic-scale_2004}.
This behavior is observed in all beam exposure experiments of STO at 200 kV: the t$_{2g}$/e$_g$ intensity ratio is the first detectable change in the fine structure, approximately at half of the electron dose compared to where changes are observed in the energy shift.
Looking at the actual Ti-L$_{2,3}$ (Fig. \ref{fig:sto_200kv_beam_exposure}e) fine structure at certain points in time in the exposure experiment (noted by arrows in c) and d)), there is a clear change consistent with the quantified values in Fig. \ref{fig:sto_200kv_beam_exposure}c and d.
All these changes are consistent with a shift from Ti$^{4+}$ to Ti$^{3+}$\cite{muller_atomic-scale_2004}.
The features quantified in the O-K core loss edge did not show any significant changes, so they are not presented here.
However, as seen in Fig. \ref{fig:sto_200kv_beam_exposure}f there are some subtle changes as a function of electron dose, mostly in peak D.

\begin{figure}[hb]
  \centering
    \includegraphics[width=1.0\columnwidth]{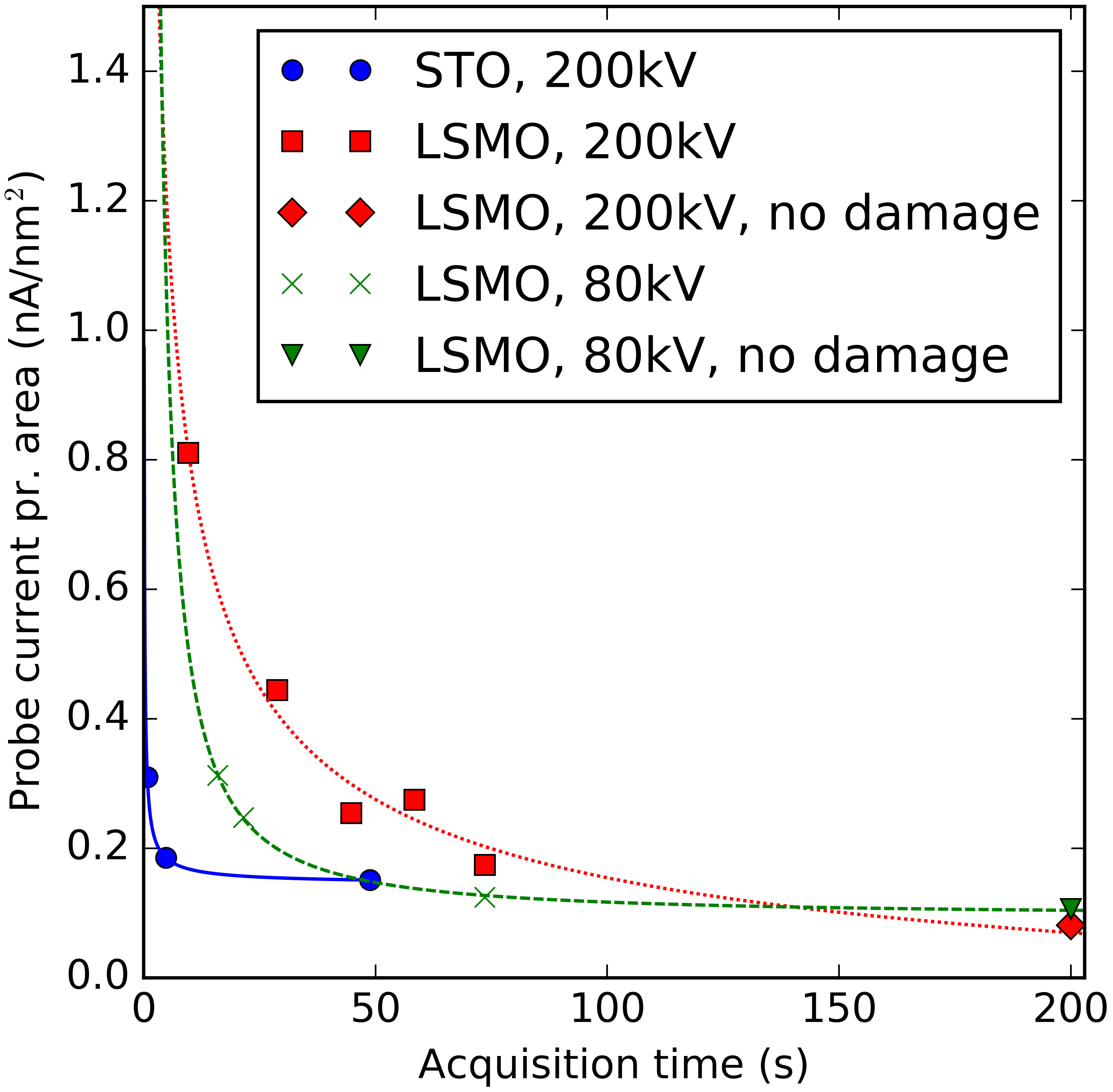}
  \caption
  {Results from several beam exposure experiments, showing the onset of damage on the material as a function of probe current per area and exposure time. 
        Green 'x' and red squares show the onset of damage on LSMO at 80 and 200 kV acceleration voltages, respectively.
        Blue circles show the same for STO at 200 kV, no beam damage was observed for STO at 80 kV.
        For sufficiently low probe current no damage was observed, even if the total electron dose was higher than the dose needed to get damage at higher probe currents.
        Some of these points are shown with the red diamond (LSMO 200 kV) and green triangle (LSMO 80 kV).
        The lines are power law fits as a guide to the eye, for showing the probe currents/acquisition times below the threshold beam damage. 
    }
    \label{fig:beam_exposure_vs_onset}
\end{figure}

The same experiment was repeated for several probe currents, which were varied by changing the condenser apertures or spot sizes.
The onset of beam damage for the different probe currents in STO and LSMO, for both 80 and 200 kV acceleration voltages, are presented in Fig. \ref{fig:beam_exposure_vs_onset}.
No beam damage was observed for STO at 80 kV, so this is not included.
Fig. \ref{fig:beam_exposure_vs_onset} shows how long one can expose the sample with a specific current before beam damage is observed.
To avoid beam damage, the combination of probe current and acquisition time must be kept below the indicated lines.
For sufficiently low probe currents no changes in the material were observed, so the data points at acquisition time = 200 s represents these.
The power law fits in Fig. \ref{fig:beam_exposure_vs_onset} are guidelines to the eye.

For all the beam exposure experiments on STO at 200 kV, the first observable changes were in the Ti-L$_{2,3}$ t$_{2g}$/e$_g$ ratio.
These changes were followed by a chemical shift of the Ti-L$_{2,3}$ peaks, and finally by a change of energy difference between the O-K prepeak and peak B.
However, for the lowest probe currents (as seen in Fig. \ref{fig:sto_200kv_beam_exposure}), the latter changes would not become visible until much later, or not at all within the exposure time.
Therefore, the best way to check for beam damage in STO at 200 kV is to look for changes in the Ti-L$_{2,3}$ t$_{2g}$/e$_g$ ratio.
For the LSMO film, fine structure changes due to beam damage eventually set in at both 80 and 200 kV acceleration voltage.
In all the experiments done with LSMO, the energy difference between the O-K prepeak and peak B was the first to show any observable changes.
In the high probe current experiments, this was followed shortly by changes in the Mn-L$_{2,3}$ intensity ratio and increasing L$_3$ and L$_2$ energy difference.
However in the low probe current experiments, no changes were observed in the Mn-L$_{2,3}$ until much later.
Thus, the best way to check for beam damage in LSMO, is to monitor for a possible energy shift of the oxygen prepeak.
Comparing the various onsets of beam damage, LSMO will be best characterized at 200 kV,  while STO should be characterized at 80 kV.
As discussed, beam damage in STO and LSMO behaves differently as a function of acceleration voltage.
For STO, there is probably a critical voltage somewhere between 80 and 200 kV, beyond which knock-on-damage sets in, since no damage was observed at 80 kV even for very high probe currents.
LSMO is more robust to knock-on-damage of the cations, but more sensitive to the introduction of oxygen vacancies\cite{yao_electron-beam-induced_2014}.

\begin{figure}[hb]
  \centering
    \includegraphics[width=0.6\columnwidth]{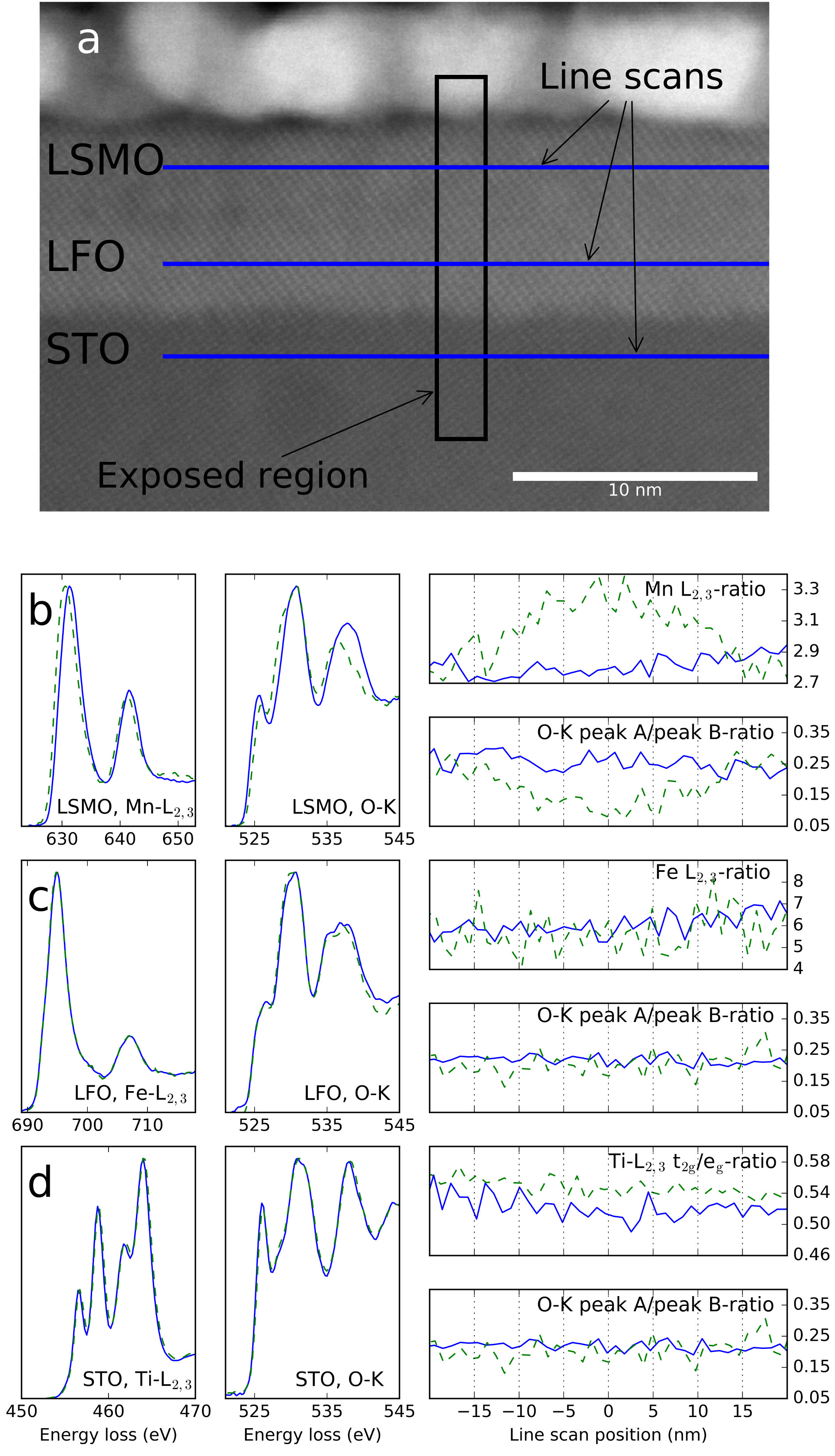}
  \caption{
      Results of an electron beam exposure experiment at 80 kV on a LSMO(4 nm)/LFO(4 nm)/STO-(111) heterostructure. 
      a) STEM-HAADF image of the region.
      Three short exposure EELS line scans were acquired parallel to the films (blue lines) before and after exposing a 2 nm wide region across the substrate and films in the form on an EELS map (black rectangle).
      Comparisons of the line scans, before and after the exposure, are shown in b), c) and d) for LSMO, LFO and STO, respectively.
      The leftmost figures show the B-cation L$_{2,3}$-edges, the middle figures the oxygen K-edge, and the rightmost figures the quantified data using EELS data processing method as explained in section \ref{sec:eels_processing_method}.
      For the LSMO shown in the b) figures, there are changes in both the Mn-L$_{2,3}$ and O-K fine structure, consistent with the electron beam damage seen in Fig. \ref{fig:stem_fine_structure_exposure}.
      The changes in the EELS fine structure are about 20 nm wide, which is much wider than the exposed region.
      The LFO and STO in c) and d), do not show any significant changes.
  }
  \label{fig:lsmo_lfo_sto_eels_map}
\end{figure}

A practical example of the effect of beam damage when doing EELS mapping is shown in Fig. \ref{fig:lsmo_lfo_sto_eels_map}.
Here, a LSMO(4 nm)/LFO(4 nm)/STO-(111) heterostructure was exposed to a large electron beam dose at 80 kV in the form of an EELS map.
Before and after the EELS map, three short exposure EELS line scans were acquired on the LSMO, LFO and STO parallel to the interface, shown in Fig. \ref{fig:lsmo_lfo_sto_eels_map}a as blue lines.
EELS data from the line scans are shown in Fig. \ref{fig:lsmo_lfo_sto_eels_map}b, c and d, for LSMO, LFO and STO, respectively.
Comparing the EELS fine structure from prior to and after the map, there are clear changes in the LSMO, but no significant changes in the LFO or STO.
The changes in the LSMO EELS fine structure are consistent with the beam damage observed in Fig. \ref{fig:stem_fine_structure_exposure}.
The electron beam induced oxygen vacancies in LSMO extend over a region (20 nm wide) that is much larger than the region (2 nm wide) exposed during acquisition of the EELS map.
This is most likely caused by oxygen diffusing from the nearby unexposed LSMO regions into the exposed region and driven by the gradient in concentration of oxygen vacancies.

When doing these kinds of long exposure acquisitions, one must first check whether the electron beam damages the material.
If damage is observed, one must try to mitigate it somehow.
As the example above shows, electron beam damage in STO can be avoided by using 80 kV.
However, since this is not a viable option for LSMO the beam dose must be reduced somehow.
The easiest way is by simply reducing the dwell time, but this will only work to a certain point due to the signal-to-noise ratio becoming too low.
One solution is to increase the spectrometer dispersion, which lead to more electron counts in each detector channel.
This gives an increase in signal-to-noise, since the effects of detector shot noise will be reduced.
On the downside, this reduces the energy resolution of the EELS data, making it harder to resolve the fine structure.
Another solution is to expose a larger area:
doing several line scans in different regions, and aligning them in post processing with respect to some feature (for example an interface), and summing them.
In the thin film systems presented in this work, this would lead to a large loss of spatial resolution in the in-plane direction, and small loss of spatial resolution in the out-of-plane direction (depending on the alignment procedure).
In practice, several of these workarounds should be combined when acquiring datasets from beam sensitive materials.

\section{Conclusions}

In the present work thresholds for electron beam damage have been determined as a function of acceleration voltage and probe current in SrTiO$_3$ and La$_{0.7}$Sr$_{0.3}$MnO$_3$. 
At 200 kV acceleration voltage, SrTiO$_3$ needs to be handled with great care to avoid beam damage that possibly can lead to misinterpretation of advanced STEM-EELS data. 
At typical probe currents used for analytical characterization, changes in both the electron energy loss fine structure and high angle annular dark field contrast are quickly observed. 
The electron beam sputters the SrTiO$_3$ TEM foil, but also changes the oxidation state of Ti as observed both by a change in the Ti-L$_{2,3}$ ratio and by a chemical shift of the Ti-L$_{2,3}$ peaks. 
Hence, high electron dose characterization of SrTiO$_3$ should better be done at a low acceleration voltage since no electron beam damage was observed at 80 kV, independent of probe current and accumulated electron dose. 

Electron beam damage occurred very differently in La$_{0.7}$Sr$_{0.3}$MnO$_3$ compared to in SrTiO$_3$. 
In La$_{0.7}$Sr$_{0.3}$MnO$_3$ the electron beam damage turned the material substoichiometric by creating oxygen vacancies. 
The loss of oxygen atoms significantly changed the oxidation state of Mn, which potentially can lead to wrong interpretations of electronic, magnetic and structural properties. 
These electron beam induced modifications of the material were observed by significant changes in the fine structure of both the O-K and the Mn-L$_{2,3}$ peaks, but not by any observable changes in any HAADF STEM images. 
Furthermore and unlike SrTiO$_3$, beam damage in La$_{0.7}$Sr$_{0.3}$MnO$_3$ occurred very similarly both at 80 and 200 kV.

\section{Acknowledgement}
The Research Council of Norway (RCN) is acknowledged for the support to the Norwegian Micro- and Nano-Fabrication Facility, NorFab (197413/V30).
Partial funding for this work was obtained from the Norwegian PhD Network on Nanotechnology for Microsystems, which is sponsored by the RCN, Division for Science, under contract no. 190086/S10.
Funding for TEM time is partly funded by the project NORTEM (Grant 197405) within the programme INFRASTRUCTURE of the RCN.
NORTEM was co-funded by the RCN and the project partners NTNU, UiO and SINTEF.
Ida Hjorth is acknowledged for fruitful discussions and proof-reading.

\section*{References}

\bibliography{bib_list}

\begin{thebibliography}{10}
\expandafter\ifx\csname url\endcsname\relax
  \def\url#1{\texttt{#1}}\fi
\expandafter\ifx\csname urlprefix\endcsname\relax\def\urlprefix{URL }\fi
\expandafter\ifx\csname href\endcsname\relax
  \def\href#1#2{#2} \def\path#1{#1}\fi

\bibitem{mannhart_oxide_2010}
J.~Mannhart, D.~G. Schlom, Oxide {Interfaces}--{An} {Opportunity} for
  {Electronics}, Science 327~(5973).
\newblock \href {http://dx.doi.org/10.1126/science.1181862}
  {\path{doi:10.1126/science.1181862}}.

\bibitem{hammerl_shedding_2011}
G.~Hammerl, N.~Spaldin, Shedding {Light} on {Oxide} {Interfaces}, Science
  332~(6032).
\newblock \href {http://dx.doi.org/10.1126/science.1206247}
  {\path{doi:10.1126/science.1206247}}.

\bibitem{martin_advances_2010}
L.~W. Martin, Y.-H. Chu, R.~Ramesh, Advances in the growth and characterization
  of magnetic, ferroelectric, and multiferroic oxide thin films, Materials
  Science and Engineering: R: Reports 68~(4-6).
\newblock \href {http://dx.doi.org/10.1016/j.mser.2010.03.001}
  {\path{doi:10.1016/j.mser.2010.03.001}}.

\bibitem{dagotto_colossal_2001}
E.~Dagotto, T.~Hotta, A.~Moreo, Colossal magnetoresistant materials: the key
  role of phase separation, Physics Reports 344~(1).
\newblock \href {http://dx.doi.org/10.1016/S0370-1573(00)00121-6}
  {\path{doi:10.1016/S0370-1573(00)00121-6}}.

\bibitem{lee_controlling_2012}
J.-S. Lee, D.~A. Arena, T.~S. Santos, C.~S. Nelson, S.~I. Hyun, J.~H. Shim,
  C.-C. Kao, Controlling competing interactions at oxide interfaces: {Enhanced}
  anisotropy in {La} 0.7 {Sr} 0.3 {MnO} 3 films via interface engineering,
  Physical Review B 85~(23).
\newblock \href {http://dx.doi.org/10.1103/PhysRevB.85.235125}
  {\path{doi:10.1103/PhysRevB.85.235125}}.

\bibitem{biegalski_impact_2015}
M.~D. Biegalski, L.~Qiao, Y.~Gu, A.~Mehta, Q.~He, Y.~Takamura, A.~Borisevich,
  L.-Q. Chen, Impact of symmetry on the ferroelectric properties of {CaTiO}3
  thin films, Applied Physics Letters 106~(16).
\newblock \href {http://dx.doi.org/10.1063/1.4918805}
  {\path{doi:10.1063/1.4918805}}.

\bibitem{he_towards_2015}
Q.~He, R.~Ishikawa, A.~R. Lupini, L.~Qiao, E.~J. Moon, O.~Ovchinnikov, S.~J.
  May, M.~D. Biegalski, A.~Y. Borisevich, Towards 3d {Mapping} of
  {BO}$_{\textrm{6}}$ {Octahedron} {Rotations} at {Perovskite}
  {Heterointerfaces}, {Unit} {Cell} by {Unit} {Cell}, ACS Nano\href
  {http://dx.doi.org/10.1021/acsnano.5b03232}
  {\path{doi:10.1021/acsnano.5b03232}}.

\bibitem{gazquez_lattice_2013}
J.~Gazquez, S.~Bose, M.~Sharma, M.~A. Torija, S.~J. Pennycook, C.~Leighton,
  M.~Varela, Lattice mismatch accommodation via oxygen vacancy ordering in
  epitaxial {La}0.5{Sr}0.5{CoO}3-$\delta$ thin films, APL Materials 1~(1).
\newblock \href {http://dx.doi.org/10.1063/1.4809547}
  {\path{doi:10.1063/1.4809547}}.

\bibitem{basletic_mapping_2008}
M.~Basletic, J.-L. Maurice, C.~Carretero, G.~Herranz, O.~Copie, M.~Bibes,
  E.~Jacquet, K.~Bouzehouane, S.~Fusil, A.~Barthelemy, Mapping the spatial
  distribution of charge carriers in {LaAlO}3/{SrTiO}3 heterostructures, Nature
  Materials 7~(8).
\newblock \href {http://dx.doi.org/10.1038/nmat2223}
  {\path{doi:10.1038/nmat2223}}.

\bibitem{khartsev_colossal_2000}
S.~I. Khartsev, P.~Johnsson, A.~M. Grishin, Colossal magnetoresistance in
  ultrathin epitaxial {La}0.75{Sr}0.25{MnO}3 films, Journal of Applied Physics
  87~(5).
\newblock \href {http://dx.doi.org/10.1063/1.372191}
  {\path{doi:10.1063/1.372191}}.

\bibitem{hallsteinsen_surface_2013}
I.~Hallsteinsen, J.~E. Boschker, M.~Nord, S.~Lee, M.~Rzchowski, P.~E. Vullum,
  J.~K. Grepstad, R.~Holmestad, C.~B. Eom, T.~Tybell, Surface stability of
  epitaxial {La}0.7{Sr}0.3{Mn}o3 thin films on (111)-oriented {SrTiO}3, Journal
  of Applied Physics 113~(18).
\newblock \href {http://dx.doi.org/10.1063/1.4804312}
  {\path{doi:10.1063/1.4804312}}.

\bibitem{oshea_nanoscale_2015}
K.~J. O'Shea, D.~A. MacLaren, D.~McGrouther, D.~Schwarzbach, M.~Jungbauer,
  S.~Huhn, V.~Moshnyaga, R.~L. Stamps, Nanoscale {Mapping} of the {Magnetic}
  {Properties} of (111)-{Oriented} {La} $_{\textrm{0.67}}$ {Sr}
  $_{\textrm{0.33}}$ {MnO} $_{\textrm{3}}$, Nano Letters\href
  {http://dx.doi.org/10.1021/acs.nanolett.5b01953}
  {\path{doi:10.1021/acs.nanolett.5b01953}}.

\bibitem{tan_2d_2011}
H.~Tan, S.~Turner, E.~Yücelen, J.~Verbeeck, G.~Van~Tendeloo, 2d {Atomic}
  {Mapping} of {Oxidation} {States} in {Transition} {Metal} {Oxides} by
  {Scanning} {Transmission} {Electron} {Microscopy} and {Electron}
  {Energy}-{Loss} {Spectroscopy}, Physical Review Letters 107~(10).
\newblock \href {http://dx.doi.org/10.1103/PhysRevLett.107.107602}
  {\path{doi:10.1103/PhysRevLett.107.107602}}.

\bibitem{kimoto_element-selective_2007}
K.~Kimoto, T.~Asaka, T.~Nagai, M.~Saito, Y.~Matsui, K.~Ishizuka,
  Element-selective imaging of atomic columns in a crystal using {STEM} and
  {EELS}, Nature 450~(7170).
\newblock \href {http://dx.doi.org/10.1038/nature06352}
  {\path{doi:10.1038/nature06352}}.

\bibitem{nakagawa_why_2006}
N.~Nakagawa, H.~Y. Hwang, D.~A. Muller, Why some interfaces cannot be sharp,
  Nature Materials 5~(3).
\newblock \href {http://dx.doi.org/10.1038/nmat1569}
  {\path{doi:10.1038/nmat1569}}.

\bibitem{muller_atomic-scale_2004}
D.~A. Muller, N.~Nakagawa, A.~Ohtomo, J.~L. Grazul, H.~Y. Hwang, Atomic-scale
  imaging of nanoengineered oxygen vacancy profiles in srtio3, Nature
  430~(7000).
\newblock \href {http://dx.doi.org/10.1038/nature02756}
  {\path{doi:10.1038/nature02756}}.

\bibitem{yao_electron-beam-induced_2014}
L.~Yao, S.~Majumdar, L.~Akaslompolo, S.~Inkinen, Q.~H. Qin, S.~van Dijken,
  Electron-{Beam}-{Induced} {Perovskite}-{Brownmillerite}-{Perovskite}
  {Structural} {Phase} {Transitions} in {Epitaxial} {La} $_{\textrm{2/3}}$ {Sr}
  $_{\textrm{1/3}}$ {MnO} $_{\textrm{3}}$ {Films}, Advanced Materials\href
  {http://dx.doi.org/10.1002/adma.201305656}
  {\path{doi:10.1002/adma.201305656}}.

\bibitem{haider_spherical-aberration-corrected_1998}
M.~Haider, H.~Rose, S.~Uhlemann, E.~Schwan, B.~Kabius, K.~Urban, A
  spherical-aberration-corrected 200kv transmission electron microscope,
  Ultramicroscopy 75~(1).
\newblock \href {http://dx.doi.org/10.1016/S0304-3991(98)00048-5}
  {\path{doi:10.1016/S0304-3991(98)00048-5}}.

\bibitem{nellist_direct_2004}
P.~D. Nellist, M.~F. Chisholm, N.~Dellby, O.~L. Krivanek, M.~F. Murfitt, Z.~S.
  Szilagyi, A.~R. Lupini, A.~Borisevich, W.~H. Sides, S.~J. Pennycook, Direct
  {Sub}-{Angstrom} {Imaging} of a {Crystal} {Lattice}, Science 305~(5691).
\newblock \href {http://dx.doi.org/10.1126/science.1100965}
  {\path{doi:10.1126/science.1100965}}.

\bibitem{sankara_rama_krishnan_misfit_2014}
P.~S. Sankara Rama~Krishnan, A.~N. Morozovska, E.~A. Eliseev, Q.~M. Ramasse,
  D.~Kepaptsoglou, W.-I. Liang, Y.-H. Chu, P.~Munroe, V.~Nagarajan, Misfit
  strain driven cation inter-diffusion across an epitaxial multiferroic thin
  film interface, Journal of Applied Physics 115~(5).
\newblock \href {http://dx.doi.org/10.1063/1.4862556}
  {\path{doi:10.1063/1.4862556}}.

\bibitem{varela_atomic-resolution_2009}
M.~Varela, M.~P. Oxley, W.~Luo, J.~Tao, M.~Watanabe, A.~R. Lupini, S.~T.
  Pantelides, S.~J. Pennycook, Atomic-resolution imaging of oxidation states in
  manganites, Physical Review B 79~(8).
\newblock \href {http://dx.doi.org/10.1103/PhysRevB.79.085117}
  {\path{doi:10.1103/PhysRevB.79.085117}}.

\bibitem{muller_atomic-scale_2008}
D.~A. Muller, L.~F. Kourkoutis, M.~Murfitt, J.~H. Song, H.~Y. Hwang, J.~Silcox,
  N.~Dellby, O.~L. Krivanek, Atomic-{Scale} {Chemical} {Imaging} of
  {Composition} and {Bonding} by {Aberration}-{Corrected} {Microscopy}, Science
  319~(5866).
\newblock \href {http://dx.doi.org/10.1126/science.1148820}
  {\path{doi:10.1126/science.1148820}}.

\bibitem{nishida_effect_2013}
S.~Nishida, S.~Kobayashi, A.~Kumamoto, H.~Ikeno, T.~Mizoguchi, I.~Tanaka,
  Y.~Ikuhara, T.~Yamamoto, Effect of local coordination of {Mn} on {Mn}-{L}2,3
  edge electron energy loss spectrum, Journal of Applied Physics 114~(5).
\newblock \href {http://dx.doi.org/10.1063/1.4817425}
  {\path{doi:10.1063/1.4817425}}.

\bibitem{egerton_radiation_2004}
R.~F. Egerton, P.~Li, M.~Malac, Radiation damage in the {TEM} and {SEM}, Micron
  35~(6).
\newblock \href {http://dx.doi.org/10.1016/j.micron.2004.02.003}
  {\path{doi:10.1016/j.micron.2004.02.003}}.

\bibitem{garvie_electron-beam-induced_1994}
L.~A.~J. Garvie, A.~J. Craven, Electron-beam-induced reduction of {Mn}4+ in
  manganese oxides as revealed by parallel {EELS}, Ultramicroscopy 54~(1).
\newblock \href {http://dx.doi.org/10.1016/0304-3991(94)90094-9}
  {\path{doi:10.1016/0304-3991(94)90094-9}}.

\bibitem{houben_spatial_2012}
L.~Houben, M.~Heidelmann, F.~Gunkel, Spatial resolution and radiation damage in
  quantitative high-resolution {STEM}-{EEL} spectroscopy in oxides, Micron
  43~(4).
\newblock \href {http://dx.doi.org/10.1016/j.micron.2011.10.006}
  {\path{doi:10.1016/j.micron.2011.10.006}}.

\bibitem{botton_elemental_2010}
G.~A. Botton, S.~Lazar, C.~Dwyer, Elemental mapping at the atomic scale using
  low accelerating voltages, Ultramicroscopy 110~(8).
\newblock \href {http://dx.doi.org/10.1016/j.ultramic.2010.03.008}
  {\path{doi:10.1016/j.ultramic.2010.03.008}}.

\bibitem{manoubi_curve_1990}
T.~Manoubi, M.~Tence, M.~G. Walls, C.~Colliex, Curve fitting methods for
  quantitative analysis in electron energy loss spectroscopy, Microscopy
  Microanalysis Microstructures 1~(1).
\newblock \href {http://dx.doi.org/10.1051/mmm:019900010102300}
  {\path{doi:10.1051/mmm:019900010102300}}.

\bibitem{francisco_de_la_pena_hyperspy:_2015}
{Francisco de la Pena}, {Pierre Burdet}, {Tomas Ostasevicius}, {Mike Sarahan},
  {Magnus Nord}, {Vidar Tonaas Fauske}, {Josh Taillon}, {Alberto Eljarrat},
  {Stefano Mazzucco}, {Gael Donval}, {Luiz Fernando Zagonel}, {Michael Walls},
  {Ilya Iyengar}, {HyperSpy} 0.8.2\href
  {http://dx.doi.org/10.5281/zenodo.28025} {\path{doi:10.5281/zenodo.28025}}.

\bibitem{ahn_inner_1985}
C.~C. Ahn, P.~Rez, Inner shell edge profiles in electron energy loss
  spectroscopy, Ultramicroscopy 17~(2).
\newblock \href {http://dx.doi.org/10.1016/0304-3991(85)90003-8}
  {\path{doi:10.1016/0304-3991(85)90003-8}}.

\bibitem{tan_oxidation_2012}
H.~Tan, J.~Verbeeck, A.~Abakumov, G.~Van~Tendeloo, Oxidation state and chemical
  shift investigation in transition metal oxides by {EELS}, Ultramicroscopy
  116.
\newblock \href {http://dx.doi.org/10.1016/j.ultramic.2012.03.002}
  {\path{doi:10.1016/j.ultramic.2012.03.002}}.

\bibitem{egerton_electron_2009}
R.~F. Egerton, Electron energy-loss spectroscopy in the {TEM}, Reports on
  Progress in Physics 72~(1).
\newblock \href {http://dx.doi.org/10.1088/0034-4885/72/1/016502}
  {\path{doi:10.1088/0034-4885/72/1/016502}}.

\bibitem{shao_quantification_2010}
Y.~Shao, C.~Maunders, D.~Rossouw, T.~Kolodiazhnyi, G.~A. Botton, Quantification
  of the {Ti} oxidation state in {BaTi}1−{xNbxO}3 compounds, Ultramicroscopy
  110~(8).
\newblock \href {http://dx.doi.org/10.1016/j.ultramic.2010.05.006}
  {\path{doi:10.1016/j.ultramic.2010.05.006}}.

\end{thebibliography}

\newpage
\begin{table}
    \caption{Properties for electron beams used in the exposure experiments.
    Probe currents were measured using the Gatan Quantum ER.}
    \label{table:probe_properties}
    \begin{tabular}{|llllll|}
        \hline
        \pbox{8cm}{Acceleration \\ Voltage} & Probe & \pbox{8cm}{Aperture \\ (\si{\micro\meter})} & \pbox{8cm}{Current \\ (pA)} & \pbox{8cm}{Size \\ (\AA)} & \pbox{8cm}{Convergence semi-\\ angle (mrad)}\\ \hline
        200 kV & 3C & 50 & 646 & 1.2 & 34.2 \\ 
        200 kV & 3C & 30 & 223 & 1.1 & 20.4 \\ 
        200 kV & 5C & 50 & 177 & 0.9 & 34.2 \\ 
        200 kV & 5C & 30 & 63 & 0.9 & 20.4 \\ \hline
        80 kV & 3C & 50 & 631 & 2.2 & 34.2 \\ 
        80 kV & 3C & 30 & 215 & 1.9 & 20.4 \\ 
        80 kV & 5C & 50 & 167 & 1.6 & 34.2 \\ 
        80 kV & 5C & 30 & 58 & 1.4 & 20.4 \\ \hline
    \end{tabular}
\end{table}

\end{document}